# A Trust-Guided Approach to MR Image Reconstruction with Side Information

Arda Atalık, Sumit Chopra, and Daniel K. Sodickson [1] [2] [3] [4]

**Abstract**

Reducing MRI scan times can improve patient care and lower healthcare costs. Many acceleration methods are designed to reconstruct diagnostic-quality images from sparse *k*-space data, via an ill-posed or ill-conditioned linear inverse problem (LIP). To address the resulting ambiguities, it is crucial to incorporate prior knowledge into the optimization problem, e.g., in the form of regularization. Another form of prior knowledge less commonly used in medical imaging is the readily available auxiliary data (a.k.a. *side information*) obtained from sources other than the current acquisition. In this paper, we present the *Trust-Guided Variational Network* (TGVN), an end-to-end deep learning framework that effectively and reliably integrates side information into LIPs. We demonstrate its effectiveness in multi-coil, multi-contrast MRI reconstruction, where incomplete or low-SNR measurements from one contrast are used as side information to reconstruct high-quality images of another contrast from heavily under-sampled data. TGVN is robust across different contrasts, anatomies, and field strengths. Compared to baselines utilizing side information, TGVN achieves superior image quality while preserving subtle pathological features even at challenging acceleration levels, drastically speeding up acquisition while minimizing hallucinations. Source code and dataset splits are available on github.com/sodicksonlab/TGVN.

Deep learning, MR image reconstruction, side information, contextual information, linear inverse problems

## Introduction

Magnetic Resonance Imaging (MRI) is a mainstay of medical diagnostic imaging, thanks to its flexibility, its rich information content, and its excellent soft-tissue contrast. An MR scanner collects measurements in frequency space (a.k.a. *k*-space) that encode the body's response to applied electromagnetic fields, typically with multiple receiver coils capturing distinct views modulated by their individual sensitivities, and this process is mathematically described by a linear map called the *forward operator*. The acquired *k*-space measurements are then used to reconstruct a spatially resolved image by solving the corresponding linear inverse problem (LIP). Despite MRI's superior diagnostic capabilities, it is comparatively time-consuming and costly, which limits its overall accessibility. Reducing the time it takes to acquire an MR scan is an important practical problem that can improve patient care by limiting patient discomfort, reducing costs, and improving accessibility of this imaging modality. One way to reduce scan time is to acquire fewer *k*-space measurements. The challenge then becomes reconstructing high-quality images from limited data by solving the corresponding ill-conditioned/ill-posed LIP, which admits many mathematically feasible solutions, most of which fail to capture essential anatomical and clinical details accurately (see Fig. 1(a)).

[1] This work was supported in part by the National Institute of Biomedical Imaging and Bioengineering (NIH P41 EB017183) and the National Science Foundation (NSF Award 1922658). Not in connection with this work, Dr. Sodickson receives fees and holds stock options as a scientific advisor for Ezra, and receives royalties from a patent on deep learning-based image reconstruction licensed by Siemens Healthineers.

[2] Arda Atalik is with the NYU Center for Data Science, and also with the Center for Advanced Imaging Innovation and Research (CAI$^2$R) and the Bernard and Irene Schwartz Center for Biomedical Imaging, Department of Radiology, NYU Grossman School of Medicine. (e-mail: Arda.Atalik@nyu.edu)

[3] Sumit Chopra is with the Courant Institute of Mathematical Sciences, and also with the Bernard and Irene Schwartz Center for Biomedical Imaging, Department of Radiology, NYU Grossman School of Medicine. (e-mail: Sumit.Chopra@nyulangone.org)

[4] Daniel K. Sodickson is with the Center for Advanced Imaging Innovation and Research (CAI$^2$R) and the Bernard and Irene Schwartz Center for Biomedical Imaging, Department of Radiology, NYU Grossman School of Medicine. (e-mail: Daniel.Sodickson@nyulangone.org)

Researchers have proposed various solutions, including compressed-sensing-based methods [2] and priors learned from exemplary data or directly from the measurements themselves [3], [4]. Recent advances in machine learning, and particularly deep learning (DL), have markedly improved the ability to tackle ill-posed or ill-conditioned problems. Notable early examples include the Variational Network (VarNet) approach [5], [6], the Model-Based Deep Learning (MoDL) approach [7], and the FISTA-Net approach [8], all of which integrate traditional optimization techniques with deep neural networks to achieve robust and efficient solutions in high-dimensional spaces. More recently, researchers have proposed generative models for reconstructing high-quality images from incomplete data [9], [10], [11], and a rapidly expanding portfolio of deep-learning-based image reconstruction methods is currently under development. In all of these cases, performance at high acceleration levels is limited by the quantity of useful information that can reliably be derived about general distributions of desirable solutions. The extent to which such general information can correctly disambiguate particular solutions is also limited. Previous work has demonstrated a sharp decline in image quality at high levels of acceleration [12].

Another approach to eliminating degenerate solutions to ill-posed or ill-conditioned LIPs involves leveraging additional contextual information (a.k.a. relevant *side information*). While side information may also be incorporated via regularizers or constraints in the objective function of an optimization problem [13], [14], [15], [16], [17], [18], it differs from population-based regularization in that it can be specific to the particular solution of interest. The nature of such side information is problem-dependent, and in many real-world scenarios it is readily available. Relevant side information can take multiple forms, including images, text, or other types of structured data. In MR image reconstruction, for instance, the side information could be data associated with prior scans of the same patient. It could also be data gathered during the same scan, such as images obtained using an imaging pulse sequence with a different underlying contrast from the target pulse sequence.[5] In more general settings, the side information need not be derived from the same imaging modality, nor does it need to be image-based; it could be textual (e.g., clinical notes and medical history), or even encoded features or representations learned from other related tasks or from foundation models.

---

[5] Note that reconstruction with different-contrast side information, also known as *conditional reconstruction*, refers to reconstructing only the target contrast while exploiting information from other contrasts. This approach differs from both single-contrast and joint multi-contrast reconstruction, though joint conditional reconstruction may also be leveraged for multi-contrast reconstruction.

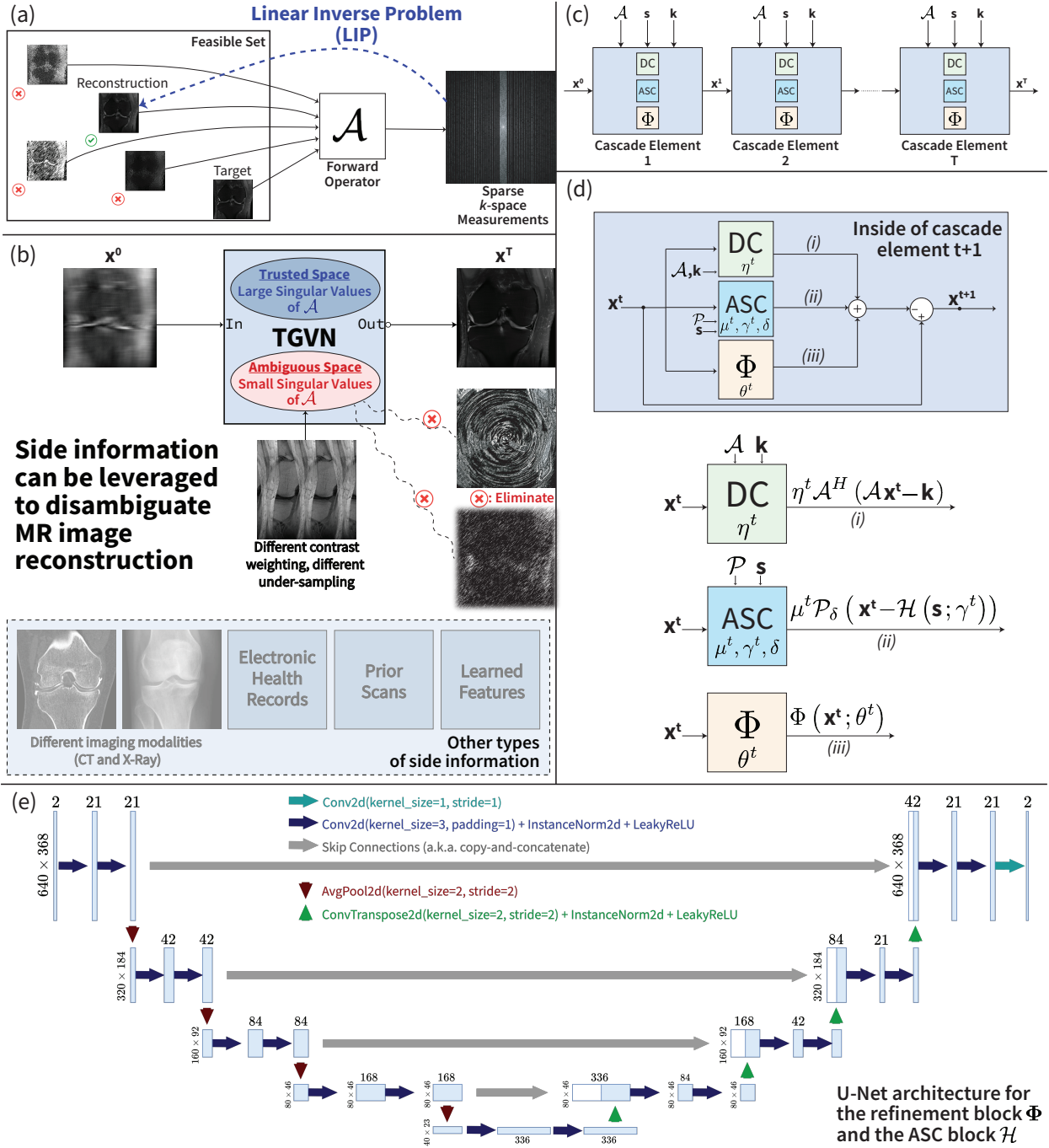

**Figure 1: *TGVN reconstruction with side information.*** *(a)* Visual representation of an LIP associated with MR image reconstruction. *(b)* Overview of trust-guided disambiguation of solutions to the LIP. In our experiments, we used different contrast-weighted measurements as side information. However, side information can originate from various sources, as highlighted within the dashed, grayed-out rectangle. *(c)* A full TGVN consisting of T cascade elements connected in series. *(d)* Components of each element in the cascade—data consistency (DC, parametrized by $\eta$), ambiguous space consistency (ASC, parametrized by $\mu, \gamma, \delta$), and refinement ($\Phi$, parametrized by $\theta$)—are shown along with their inputs, their outputs, and the final aggregation of outputs. *(e)* U-Net [1] architecture used in the refinement block $\Phi$ and the ASC block $\mathcal{H}$, illustrated for a

$640 \times 368$ *coil-combined image, with 4 pooling levels and an initial* 21 *channels in the first double-convolution, doubling at each pooling level.*

## Contributions

In this work, we propose a novel, end-to-end trainable deep learning method that efficiently and reliably integrates side information to solve LIPs, as illustrated in Fig. 1. Our method, called the **T**rust-**G**uided **V**ariational **N**etwork (**TGVN**), uses the side information only to disambiguate the subspace of solutions that the forward operator cannot reliably distinguish based on the measured data (i.e., the *ambiguous space*). Specifically, we introduce a learnable squared Euclidean distance constraint, termed the *ambiguous space consistency* constraint, into the regularized least-squares reconstruction formulation of the LIP to eliminate undesirable solutions from the ambiguous space of the forward operator. This ambiguous space consistency constraint can be seamlessly integrated into any deep unrolled network. Our approach can be trained end-to-end with full supervision to maximize a similarity metric between the reconstructed and the ground-truth image, requiring minimal modifications to integrate the constraint. We demonstrate the effectiveness of our method in the challenging domain of multi-coil, multi-contrast MR image reconstruction, where incomplete or low-SNR measurements from complementary contrast weighting are used as side information to reconstruct images with a different target contrast from exceedingly small quantities of *k*-space measurements (on the order of $20 \times$ undersampling in a single phase-encoding direction) across different anatomies and field strengths. Compared to recently proposed DL-based solutions, our method leverages side information more efficiently and reliably, preserving fine anatomical and pathological details obliterated by other methods at high acceleration levels, and achieving statistically significant improvements in reconstruction performance while also being robust to degradations in the quality of side information. To summarize:

- We propose the novel TGVN method leveraging side information to reliably solve ill-posed or ill-conditioned LIPs.

- We demonstrate the effectiveness of TGVN in multi-coil multi-contrast MR image reconstruction, using incomplete or low-SNR measurements from complementary contrast weighting as side information.

- We demonstrate the robustness of our method by showing its efficacy for different contrasts across multiple anatomies and multiple field strengths. We also report the results of various ablation studies to illuminate the origins of this robustness.

- We show that TGVN leverages side information more efficiently than other recent ML-based solutions, achieving statistically significant improvements in image reconstruction performance and pushing the boundaries of current techniques in medical imaging and beyond.

# Background

## Multi-Coil MR Image Acquisition

In MR imaging, measurements are acquired in the spatial frequency (a.k.a. *k*-space) domain, and the measurements are related to the estimated MR image through the linear forward operator $\mathcal{A}$.

These measurements may be grouped into a complex-valued vector $\tilde{\mathbf{k}}$, and the elements of $\tilde{\mathbf{k}}$ represent Fourier coefficients of the structure of the continuous object being imaged. Specifically, we define a discrete estimated MR image $\mathbf{x}$, such that $\tilde{\mathbf{k}} = \mathcal{F}(\mathbf{x}) + \epsilon$, where $\epsilon$ is complex Gaussian noise and $\mathcal{F}$ denotes the Fourier transform operator. The vector $\mathbf{x} \in \mathbb{C}^{MN}$ is a complex vector of size $MN$, where $M$ and $N$ are pixel dimensions of the two-dimensional image being sought.

In parallel imaging (PI), the scanner captures multiple views of the anatomy modulated by the sensitivities $S_i$ of the receiver coils, which can be represented by diagonal matrices $S_i \in \mathbb{C}^{MN \times MN}$. In this case the relationship becomes $\tilde{\mathbf{k}}_i = \mathcal{F}(S_i \mathbf{x}) + \epsilon_i$, for each $i \in \{1, 2 \ldots, N_c\}$, where $N_c$ denotes the number of coils. To simplify notation, we aggregate the *k*-space data from all coils into a single tensor $\tilde{\mathbf{k}} = (\tilde{\mathbf{k}}_1, \ldots, \tilde{\mathbf{k}}_{N_c})$ and define the *expand* operator ($\mathcal{E}$) which maps the complex image to multi-coil *k*-space. That is, $\mathcal{E}: \mathbf{x} \mapsto \left( \mathcal{F}(S_1 \mathbf{x}), \ldots, \mathcal{F}(S_{N_c} \mathbf{x}) \right)$. To accelerate MR acquisition, fewer *k*-space samples are acquired, which we denote by a binary diagonal mask $\mathcal{M} \in \{0,1\}^{MN \times MN}$. Thus, the set of under-sampled *k*-space measurements can be denoted as $\mathbf{k} \triangleq \mathcal{M}\tilde{\mathbf{k}} = (\mathcal{M}\tilde{\mathbf{k}}_1, \ldots, \mathcal{M}\tilde{\mathbf{k}}_{N_c})$, and the forward operator $\mathcal{A}$ mapping the underlying image to the under-sampled and noisy *k*-space measurements in multi-coil MR image acquisition is equal to $\mathcal{M} \circ \mathcal{E}$. That is,

$$\mathbf{k} = \mathcal{A}\mathbf{x} + \epsilon' = (\mathcal{M} \circ \mathcal{E})\mathbf{x} + \epsilon', \tag{1}$$

where $\epsilon'$ denotes the complex Gaussian noise in the under-sampled *k*-space measurements.

## Deep Learning for Parallel MR Image Reconstruction

Given the forward operator $\mathcal{A}$ and the *k*-space data $\mathbf{k}$, estimating $\mathbf{x}$ is considered a *well-posed* problem if it meets the following three criteria (called the Hadamard conditions): 1) existence of a solution, 2) uniqueness of the solution, and 3) stability of the solution [19]. Accelerated parallel MR image reconstruction, however, like most real-word problems, is either *ill-posed*, failing to meet one or more of these criteria, or *ill-conditioned*, with small errors in the measurements leading to much larger errors in our image estimate $\mathbf{x}$. This is because the sparse set of measurements $\mathbf{k}$ makes the above system of equations (1) either underdetermined, with a potentially infinite set of solutions, or ill-conditioned, with a large yet finite condition number. When the measurement noise is Gaussian, the maximum likelihood estimate of a solution to (1) is given by $\hat{\mathbf{x}} = \arg\min_{\mathbf{x}} \frac{1}{2} \|\mathcal{A}\mathbf{x} - \mathbf{k}\|_2^2$. To address its ill-posed or ill-conditioned nature, the LIP is reformulated to impose additional constraints or requirements on the solution. By incorporating appropriate additional constraints, one can derive a reliable approximate solution. More formally, let $\Psi(\cdot)$ denote a regularization function that imposes certain constraints on the possible solutions $\mathbf{x}$, e.g., sparsity in the wavelet or total variation domain. Then, the optimization problem can be modified as follows:

$$\hat{\mathbf{x}} = \arg\min_{\mathbf{x}} \frac{1}{2} \|\mathcal{A}\mathbf{x} - \mathbf{k}\|_2^2 + \Psi(\mathbf{x}). \tag{2}$$

In deep-learning based unrolled networks, such as the End-to-end Variational Network (E2E-VarNet) [6], one learns a regularization function from the training data to maximize a desired similarity metric between the reconstructed image $\hat{\mathbf{x}}$ and the ground truth. Specifically, E2E-VarNet starts with an initial estimate $\mathbf{x}^0$ of the solution to $\mathcal{A}\mathbf{x} = \mathbf{k}$, and uses the Landweber method [20] to iteratively refine its estimate. Furthermore, it replaces the gradient of the regularization function $\Psi(\mathbf{x})$ with a neural network $\Phi$, parametrized by $\theta^t$ at each iteration $t$. More formally, E2E-VarNet executes the following sequence of steps for a total of $T$ iterations (implemented in $T$ cascade elements similar to Fig. 1), starting with $\mathbf{x}^0 = \mathcal{A}^H \mathbf{k}$:

$$\mathbf{x}^{t+1} = \mathbf{x}^t - \eta^t \mathcal{A}^H (\mathcal{A} \mathbf{x}^t - \mathbf{k}) - \Phi(\mathbf{x}^t; \theta^t), \tag{3}$$

where $\mathcal{A}^H = \mathcal{E}^H \circ \mathcal{M}$ is the Hermitian adjoint of $\mathcal{A}$. It is worth mentioning that the second term on the right hand side is usually referred to as *data consistency*, as it guides $\mathbf{x}$ to be maximally consistent with the acquired measurements. At the end of iteration $T$, we obtain $\mathbf{x}^T$ parameterized by $\Theta = \{\theta^0, \ldots, \theta^{T-1}, \eta^0, \ldots, \eta^{T-1}\}$. Assuming access to ground truth $\mathbf{x}^*$, parameters Θ are learned in a supervised manner to maximize a desired similarity between $\mathbf{x}^T$ and $\mathbf{x}^*$.

## Related Work

We outline how prior work has utilized side information in MR image reconstruction. While side information can take various forms, most studies have focused on complementary contrast information—reconstructing the target contrast by leveraging information from other contrast(s). As was mentioned earlier, this approach differs from both single-contrast and joint multi-contrast reconstruction.

### Initial Attempts

The use of side information in medical image reconstruction dates back to at least the 1990s. [21] demonstrated tomographic image reconstruction based on a weighted Gibbs penalty, where the weights are determined by anatomical boundaries in high-resolution MR images. [22] proposed a Bayesian method whereby maximum a posteriori (MAP) estimates of PET and SPECT images may be reconstructed with the aid of prior information derived from registered anatomical MR images of the same slice. Some of the earlier attempts also utilized handcrafted priors [13], [14], [15], [23], [24], [25], [26], [27], [28], [29]. Later, dictionary-learning-based methods were proposed [16], [30].

### End-to-End Methods

More recently, multiple authors have proposed end-to-end deep learning-based models that leverage side information for MR image reconstruction. Specifically, [31], [32] proposed combining T1-weighted images and under-sampled T2-weighted images to reconstruct fully sampled T2-weighted images using a Dense U-Net model. [17] introduced a Dilated Residual Dense Network (DuDoRNet) for dual domain restorations from under-sampled MRI data to simultaneously recover *k*-space and images. [33] developed a multi-modal transformer ('MTrans') for accelerated MR imaging which transferred multi-scale features from the target modality to the auxiliary modality. Rather than manually designing fusion rules, [18] presented a multi-contrast VarNet ('MC-VarNet') to explicitly model the relationship between different contrasts.

### Generative Models

Generative models utilizing side information for MR image reconstruction are GAN-based and score-based algorithms. These models can be divided into reconstruction and synthesis methods, in which the former is our focus. Specifically, [34] utilized conditional GANs with three priors—shared high-frequency, low-frequency, and perceptual priors. [35] proposed a framework for estimating objects from incomplete imaging measurements by optimizing in the latent space of a style-based generative model, using constraints from a related prior image. [36] introduced a score-based generative model ('DMSI') to learn a joint Bayesian prior over multi-contrast data.

## Range–Null Space Decomposition

Range–null space decomposition involves breaking down a vector space into two orthogonal subspaces: the range (or column space) and the null space of a linear operator. Given a linear operator $\mathcal{A}$, the range space consists of all possible outputs of $\mathcal{A}$, while the null space contains all vectors that are mapped to zero by $\mathcal{A}$. This decomposition is particularly useful in solving LIPs, as it allows for the separation of components that are preserved by the operator from those that are not. The combination of range–null space decomposition with deep learning was first explored in [37]. More recently, [38] introduced a GAN-based super-resolution method that learns only the null-space component while fixing the range-space component; [39] presented a diffusion model for LIPs which refines only the null-space contents during the reverse diffusion process; and [40] addressed the inconsistencies in MR image reconstruction by decomposing input data and selectively feeding the null-space component into proximal mapping.

Despite notable advancements, existing methods that incorporate side information for solving LIPs still struggle with highly under-sampled data, often producing reconstructions with degraded image quality or hallucinations. The former can be attributed to a lack of efficiency in exploiting side information (i.e., insufficient ability to disambiguate the solution space), while the latter represents over-reliance on it. Consequently, harnessing the full potential of side information while mitigating hallucinations remains an open problem that can have a transformative impact on the efficiency and accessibility of medical imaging.

## Trust Guided Variational Network (TGVN)

We now give details of our proposed method, that effectively and reliably leverages *side information* to impose additional constraints into the LIP and guide the solution to fall within a contextually appropriate distribution. In this setting, we assume that we have access to the additional side information denoted by $\mathbf{s}$ when solving for $\mathbf{x}$ using the system of equations $\mathcal{A}\mathbf{x} = \mathbf{k}$. So long as $\mathbf{s}$ and $\mathbf{x}$ are conditionally dependent given $\mathbf{k}$ (i.e., the conditional mutual information $I(\mathbf{s};\mathbf{x}|\mathbf{k}) > 0$), the knowledge of $\mathbf{s}$ can be exploited to reduce the uncertainty in estimating $\mathbf{x}$ from $\mathbf{k}$ [41]. As such, our solution assumes the existence of such conditional dependence.

### The Motivation: Ambiguous Space Consistency

Deep learning and physics-based unrolled networks have shown notable success in MR image reconstruction from sparse *k*-space data [42], primarily due to their ability to enforce *data consistency*—ensuring that the reconstructed images closely match the acquired measurements. However, while data consistency is crucial for aligning the solution with the observed data, it might not be sufficient to resolve inherent ambiguities in the solution space, particularly at higher accelerations where an abrupt degradation in image quality has been highlighted [12], rendering the images non-diagnostic. To address this issue, we introduce the concept of *ambiguous space consistency*, which goes beyond data consistency and complements it. Essentially, our idea is to identify a subspace of images that could significantly alter reconstruction quality without substantially affecting the data inconsistency loss—$\|\mathcal{A}\mathbf{x} - \mathbf{k}\|_2^2$— of the MR image reconstruction problem. We then aim to use side information preferentially in this ambiguous space to resolve ambiguities without compromising information that has been encoded reliably in measured data. Conceptually, one might expect the ambiguous space to be associated with low singular values of the forward problem, since it is known that singular values are largest in the reverse mapping where they were smallest in the forward mapping, resulting in amplified noise and increased sensitivity to small perturbations.

Let $\mathbf{x}_p$ be a particular solution to the equation $\mathcal{A}\mathbf{x} = \mathbf{k}$ and let $\sum_i \sigma_i \mathbf{u}_i \mathbf{v}_i^H$ denote the singular value decomposition (SVD) of $\mathcal{A}$. Given a small positive threshold $\delta$, we define the *ambiguous space* as the subspace spanned by the right singular vectors $\mathbf{v}_i$ with corresponding singular values $\sigma_i$ smaller than $\delta$, and denote it as $\mathcal{W}_\delta(\mathcal{A})$. Observe that if we add any unit vector $\mathbf{x}_a \in \mathcal{W}_\delta(\mathcal{A})$ to $\mathbf{x}_p$, the data inconsistency loss $\|\mathcal{A}(\mathbf{x}_p + \mathbf{x}_a) - \mathbf{k}\|_2^2$ can at most be $\delta^2$. In other words, perturbing a solution that aligns with the observed measurements by adding a vector from the ambiguous space results in only a minor change to the objective value. Perturbation in the ambiguous space can create only minor data inconsistency. However, only certain $\mathbf{x}_a$ maximize the desired similarity between $\mathbf{x}_p + \mathbf{x}_a$ and $\mathbf{x}^*$, indicating that, once a particular solution is found, images from $\mathcal{W}_\delta(\mathcal{A})$ introduce ambiguity in the reconstruction problem. That is, they might visually alter the reconstruction quality without significantly affecting the data inconsistency loss $\|\mathcal{A}\mathbf{x} - \mathbf{k}\|_2^2$. Inspired by this observation, we propose to explicitly learn a constraint that removes undesirable solutions from $\mathcal{W}_\delta(\mathcal{A})$. Our idea is to project $\mathbf{x}$ onto $\mathcal{W}_\delta(\mathcal{A})$ with the orthogonal projector $\mathcal{P}_\delta$ to obtain $\mathbf{x}_a$, and to guide $\mathbf{x}_a$ to be maximally consistent with the side information $\mathbf{s}$ using a learnable module $\mathcal{H}$ parametrized by $\gamma$. Specifically, we add a squared Euclidean distance constraint $\|\mathcal{P}_\delta \mathbf{x} - \mathcal{H}(\mathbf{s}; \gamma)\|_2^2$ to (2) to obtain

$$\hat{\mathbf{x}} = \arg\min_{\mathbf{x}} \frac{1}{2}\|\mathcal{A}\mathbf{x} - \mathbf{k}\|_2^2 + \frac{\beta}{2}\|\mathcal{P}_\delta \mathbf{x} - \mathcal{H}(\mathbf{s}; \gamma)\|_2^2 + \Psi(\mathbf{x}). \tag{4}$$

Ambiguous space consistency (ASC) is both analogous to and complementary to data consistency (DC). Both are linear in $\mathbf{x}$, and both ensure that a linear transformation applied to $\mathbf{x}$ is close to a desired vector—acquired measurements in DC and a learned transformation of side information in ASC. This linearity helps to maintain the simplicity of Landweber iterations, since the gradient of the ASC with respect to $\mathbf{x}$ is simply $\mathcal{P}_\delta(\mathbf{x} - \mathcal{H}(\mathbf{s}; \gamma))$. By design, however, ASC operates in the ambiguous space. As a result, it does not adversely affect DC, thereby minimizing the risk of hallucinations resulting from over-reliance on side information. This is a crucial distinction between our approach to incorporating side information and other methods.

**Table 1**: *Experiments* – Undersampling, in either a random or a uniform pattern, was performed along the phase-encoding (PE) direction. Center frequency (CF) is the portion of fully sampled central PE lines. For knee experiments (K1, K2, and K3), acceleration was achieved solely by undersampling. For brain experiments (B1 and B2), the total acceleration factors include both undersampling (18 × and 15 × for main information) and reduction in the number of repetitions (2 × and 3 × for main, 3 × for side information).

| Experiment | Contrast | Acceleration | CF | Contrast | Acceleration | CF |
|---|---|---|---|---|---|---|
| K1 | PDFSw | Random-20 × | 3% | PDw | Uniform-2 × | 0% |
| K2 | PDFSw | Random-14 × | 3% | PDw | Uniform-3 × | 0% |
| K3 | PDFSw | Random-6 × | 5% | PDw | Uniform-3 × | 0% |
| B1 | FLAIR | Uniform-36 × | 2% | T2w | 3 × | – |
| B2 | T1w | Uniform-45 × | 2% | T2w | 3 × | – |

Our reason for choosing a more general projector $\mathcal{P}_\delta$ rather than simply using an orthogonal projector onto the null space of $\mathcal{A}$ is twofold. First, in practice, the forward operator matrix $\mathcal{A}$ in parallel MR imaging and other LIPs can have a trivial null space but still exhibit many small, non-zero singular values. This is the reason for high noise amplification at higher acceleration rates [43]. Second, even when the null space is non-trivial (i.e., it does not only contain the zero vector), the presence of small singular values can pose challenges, and the proposed approach can further assist in resolving these ambiguities. In the limit as $\delta$ approaches 0 from above, the set $\mathcal{W}_\delta(\mathcal{A})$ converges to the null space of $\mathcal{A}$ (i.e., $\lim_{\delta \downarrow 0} \mathcal{W}_\delta(\mathcal{A}) = \mathcal{N}(\mathcal{A})$) and the projection $\mathcal{P}_\delta$ approaches the orthogonal projector onto $\mathcal{N}(\mathcal{A})$. Hence, TGVN is more general than simply resolving the null space—it naturally encompasses null space resolution as a special case.

## The Solution: Iterative Optimization

The solution to (4) can be obtained using a cascade of neural networks similar to those used in the E2E-VarNet method. As the added constraint involves only a squared Euclidean distance, its integration into (3) is straightforward. Starting with $\mathbf{x}_0 = \mathcal{A}^H \mathbf{k}$, we execute the following sequence of steps for a total of $T$ iterations.

$$\mathbf{x}^{t+1} = \mathbf{x}^t - \eta^t \mathcal{A}^H(\mathcal{A}\mathbf{x}^t - \mathbf{k}) - \underbrace{\mu^t \mathcal{P}_\delta\left(\mathbf{x}^t - \mathcal{H}(\mathbf{s}; \gamma^t)\right)}_{\text{trust-guidance}} - \Phi(\mathbf{x}^t; \theta^t). \tag{5}$$

At the end of iteration $T$, we obtain $\mathbf{x}^T$ parameterized by $\Omega \triangleq \Theta \cup \{\delta, \gamma^0, \dots, \gamma^{T-1}, \mu^0, \dots, \mu^{T-1}\}$. Assuming access to ground truth $\mathbf{x}^*$, the parameters $\Omega$ are learned in a supervised manner to maximize a desired similarity between $\mathbf{x}^T$ and $\mathbf{x}^*$. It is worth noting that the parameter $\delta$ can be learned from the data as proposed, or it can be fixed based on the coil specifications and undersampling pattern by analyzing the distribution of singular values.

In high-dimensional problems like parallel MR imaging, the computational burden of working directly with large-scale operators can be prohibitive. Therefore, instead of explicitly calculating the SVD of the forward operator, which would be computationally expensive, we seek an efficient alternative. Here, we present an efficient approximation of the exact orthogonal projector $\mathcal{P}_\delta$, which bypasses the need for SVD computation. This approach is crucial for managing the scale of the forward operator, which may contain hundreds of thousands of rows and columns, making explicit methods infeasible. For a set $\mathcal{K}$, let $1_\mathcal{K}(x)$ denote an indicator function that equals 1 if $x \in \mathcal{K}$ and 0 otherwise. Given $\delta$, the exact projector can be written as $\mathcal{P}_\delta = \sum_i 1_{[0,\delta)}(\sigma_i) \mathbf{v}_i \mathbf{v}_i^H$. In lieu of assigning binary weights to the $i$th projection, we can weight them by $\delta^2/(\delta^2 + \sigma_i^2)$, and define

$$\mathcal{P}'_\delta \triangleq \sum_i \frac{\delta^2}{\delta^2 + \sigma_i^2} \mathbf{v}_i \mathbf{v}_i^H = \left(I + \frac{1}{\delta^2} \mathcal{A}^H \mathcal{A}\right)^{-1}, \tag{6}$$

where $I: \mathbf{x} \mapsto \mathbf{x}$ denotes the identity operator. We can then calculate the approximate trust-guidance term

$$\mu^t \left(I + \frac{1}{\delta^2} \mathcal{A}^H \mathcal{A}\right)^{-1} \left(\mathbf{x}^t - \mathcal{H}(\mathbf{s}; \gamma^t)\right), \tag{7}$$

efficiently using a small number of Conjugate Gradient iterations [44].

One can build intuition about ambiguous space consistency by considering a simplified example. For the special case of a single-coil acquisition with a binary undersampling mask $\mathcal{M}$, the approximate projector onto the ambiguous space is given by $\mathcal{P}'_\delta = \mathcal{F}^H \left(\frac{\mathcal{M}}{\delta^2} + I\right)^{-1} \mathcal{F}$. Here, $\mathcal{P}'_\delta \mathbf{x}$ first transforms the complex-valued image **x** into *k*-space. In the *k*-space domain, the acquired lines are scaled by $\frac{\delta^2}{1+\delta^2} \ll 1$, while the non-acquired lines retain unit weight. The weighted *k*-space is then transformed back into image space by $\mathcal{F}^H$. Since $\delta$ is small, the acquired lines are significantly attenuated, causing the non-acquired (masked) lines to dominate the projection. For $\delta = \frac{1}{3}$, for example, acquired lines are scaled by a factor of 0.1, while the masked lines remain unscaled. In this case, side information contributes principally to the non-acquired *k*-space lines while still retaining a nonzero contribution to the acquired lines. In the limit $\delta \downarrow 0$, side information contributes only to the non-acquired *k*-space lines. It should be noted that in the multi-coil setting, coil sensitivities play a role in spatial encoding, and the distinction between acquired and non-acquired lines is less clear, but the same general principles apply to the separation of trusted from untrusted components.

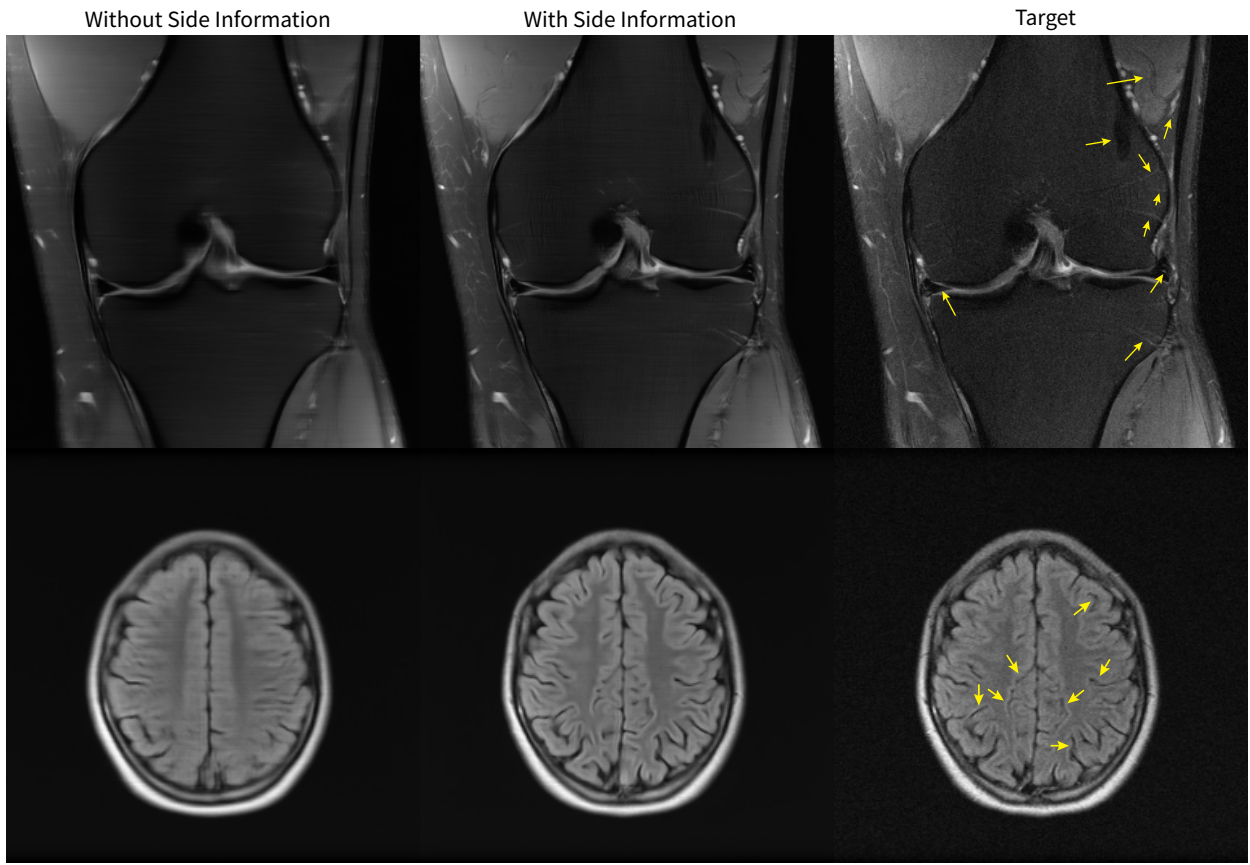

**Figure 2:** *Leveraging side information significantly enhances the reconstruction quality. **Top:** Coronal PDFS knee image reconstructions at $20 \times$ acceleration compared with unaccelerated reference image. **Bottom:** Axial FLAIR brain image reconstructions at $36 \times$ acceleration compared with unaccelerated reference image. **Left:** Reconstructed MR image from highly sparse MR measurements using E2E-VarNet. **Middle:** Reconstructed MR image from the same sparse MR measurements, with additional side information from a different sequence using TGVN (having the same capacity as the E2E-VarNet). **Right:** Ground-truth target image, with prominent anatomical features highlighted by yellow arrows. These features—such as subtle intrameniscal signals and the vastus lateralis muscle in the top figure, and fine neuroanatomical details in the bottom figure—are preserved only when side information is used.*

## Empirical Validation

We validated the efficacy of TGVN by using it for multi-coil MR image reconstruction from different contrasts across different anatomies and field strengths, as detailed in Table 1. In all experiments, we utilized the efficient approximate projection introduced in (6). In our empirical validation, we sought answers to the following four questions:

Q1. Is there any benefit in using the side information?

Q2. How effective is TGVN at utilizing the side information?

Q3. Does projecting onto the *ambiguous space* provide any benefits compared to no projection?

Q4. How robust is the proposed approach to different undersampling factors, to misregistration between images used for main and side information, and to degradation in the quality of side information?

To answer Q1, we compared the reconstruction performance of TGVN against an E2E-VarNet of the same capacity that does not utilize side information. Q2 was answered by comparing the performance of TGVN against several recent DL baselines that also leverage side information in image reconstruction: MTrans [33], MC-VarNet [18], and DMSI [36]. To address Q3, we compared the performance of TGVN with and without the projection. Q4 was answered by conducting experiments using multiple undersampling factors in main and side information, by deliberately introducing misregistrations, and by using irrelevant side information in the form of random samples from a Gaussian distribution. We present our findings related to the first and second questions in Secs. 5.1 and 5.2. Our findings related to the third and fourth questions are presented in Sec. 5.3. In our experiments, undersampling for both the main and side information was implemented along the phase-encoding direction. The target images were selected as the root-sum-of-squares (RSS) combination $\sqrt{\sum_i |\mathbf{x}_i|^2}$ of fully sampled component coil images $\mathbf{x}_i$. We evaluated the reconstruction quality using three metrics: the structural similarity index [45] (SSIM, in %), peak signal-to-noise ratio (PSNR, in dB), and normalized root-mean-squared error (NRMSE). Additional training and evaluation details are provided in the Appendix 6.1. To demonstrate the statistical significance of the improvements in image reconstruction metrics, we performed a Wilcoxon signed-rank test [46] between the metrics calculated on the test dataset for TGVN ($s_{\text{ours}}$) as compared to the best-performing baseline ($s_{\text{base}}$, the baseline with the best average score).

In knee experiments (K1, K2, K3), we utilized a subset of the multi-coil track of the fastMRI knee dataset—an open-source dataset consisting of *k*-space measurements from clinical 3T and 1.5T scanners paired with the ground-truth clinical cross-sectional images [47]. Our dataset comprised coronal MR scans of 428 patients using a proton-density weighting with fat suppression (PDFSw) and proton density weighting without fat suppression (PDw). Data acquisition employed a 15-channel knee coil array and Cartesian 2D Turbo Spin Echo (TSE) pulse sequences. We split the dataset into three non-overlapping subsets of sizes 368, 30, and 30 image volumes, for training, validation, and test sets, respectively, with a total of 15,231 slices. It should be noted that pathology labels were available for this dataset [48]; even the most prevalent pathology—meniscus tear—appears in only 2,163 out of 15,231 slices, highlighting the sparsity of pathological features at the slice level. We visually assessed reconstruction performance by comparing slices exhibiting pathology (verified by an experienced musculoskeletal radiologist) during inference.

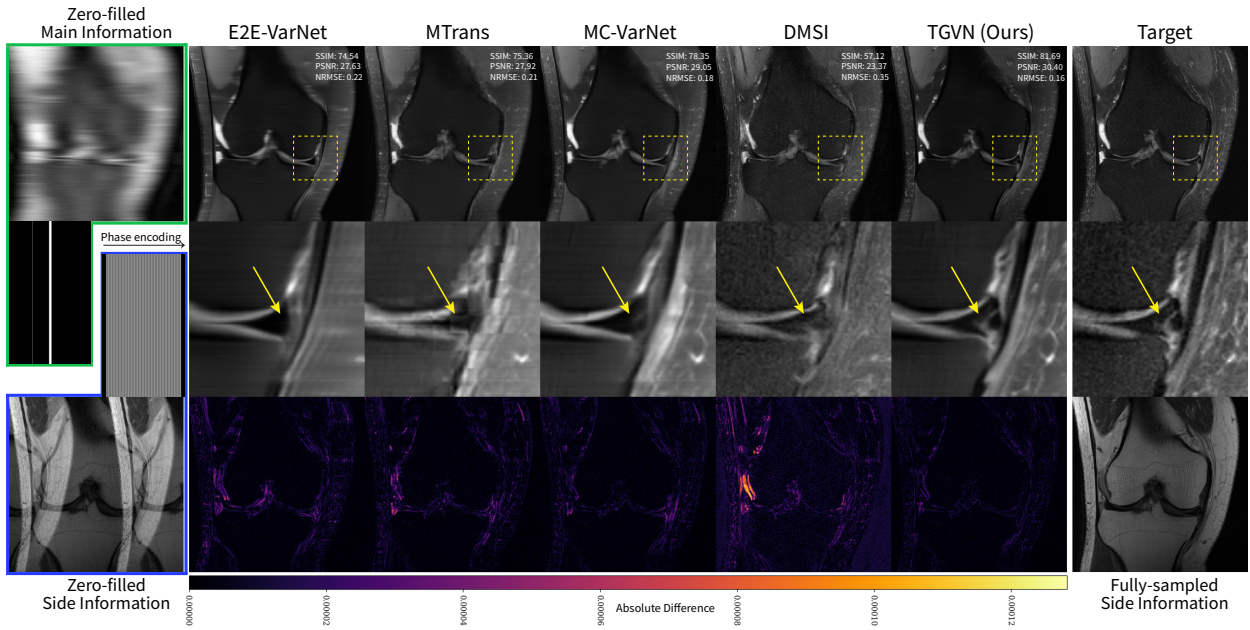

**Figure 3:** *Knee image reconstructions from K1 showing the effectiveness of TGVN in leveraging side information.* TGVN is able to reconstruct a high-quality image even at challenging undersampling levels of $20\times$, in comparison to various baselines. The medial meniscus tear (yellow arrow) is clearly visible only in the TGVN reconstruction, despite being obscured in the fully sampled side information. **Top:** *Full field of view images.* **Middle:** *Undersampling masks for main and side information (left), along with zoomed-in regions indicated by dashed yellow boxes in the top row.* **Bottom:** *Zero-filled reconstruction of under-sampled side information used as input (left), absolute difference maps (center) between the target image and various reconstructions, using a consistent color map shown at bottom, and a fully sampled image of the side information (right, shown for illustration only. Models only had access to $2\times$ under-sampled side information in K1). TGVN exhibits the smallest error and the best reconstruction metrics.*

Brain experiments (B1, B2) utilized the M4Raw dataset [49]—a publicly available multi-channel *k*-space dataset of brain scans of 183 healthy volunteers acquired using a low-field (0.3T) scanner. It includes axial MR scans with three contrasts, acquired using a 4-channel array: T1-weighted (T1w), T2-weighted (T2w), and fluid-attenuated inversion recovery (FLAIR). Each scan has 18 slices per contrast with varying numbers of repetitions. We used single-repetition measurements to reconstruct images with quality similar to that of multi-repetition aggregated RSS targets. The training, validation, and test sets included 128, 30, and 25 volumes, respectively. In both knee and brain experiments, we ensured that patient-level and study-level splits were used, so that any single patient's data appears exclusively in the training, the validation, or the test set only. Details regarding experimental settings, including contrast and acceleration details, are provided in Table 1.

**Table 2: *Quantitative evaluation results*** – SSIM, PSNR, and NRMSE are shown for knee (K1, K2, K3) and brain (B1, B2) experiments using TGVN and baselines. For each evaluation metric and each reconstruction method, the mean and standard error of the mean over the test dataset are reported. Bold-face values indicate the best performance in each category, which in all cases is achieved by TGVN reconstruction. For SSIM and PSNR, higher is better; for NRMSE, lower is better.

| Metric | Exp. | TGVN | DMSI | MC-VarNet | MTrans | E2E-VarNet |
|---|---|---|---|---|---|---|
| SSIM | K1 | **84.92 ± 0.19** | 56.99 ± 0.31 | 82.89 ± 0.21 | 80.84 ± 0.23 | 81.33 ± 0.23 |
| | K2 | **85.52 ± 0.19** | 58.76 ± 0.31 | 83.13 ± 0.21 | 81.25 ± 0.22 | 83.40 ± 0.21 |
| | K3 | **88.02 ± 0.17** | 64.35 ± 0.31 | 86.47 ± 0.18 | 85.29 ± 0.19 | 87.42 ± 0.17 |
| | B1 | **87.34 ± 0.12** | - | 86.95 ± 0.12 | 84.03 ± 0.14 | 81.40 ± 0.15 |
| | B2 | **89.34 ± 0.10** | - | 88.66 ± 0.12 | 85.72 ± 0.17 | 86.11 ± 0.13 |
| PSNR | K1 | **30.92 ± 0.07** | 22.22 ± 0.10 | 29.97 ± 0.07 | 28.93 ± 0.07 | 29.30 ± 0.07 |
| | K2 | **31.31 ± 0.07** | 22.68 ± 0.10 | 30.07 ± 0.07 | 29.11 ± 0.07 | 30.37 ± 0.07 |
| | K3 | **32.89 ± 0.08** | 24.24 ± 0.10 | 31.95 ± 0.07 | 31.26 ± 0.07 | 32.59 ± 0.07 |
| | B1 | **30.81 ± 0.08** | - | **30.75 ± 0.08** | 28.70 ± 0.08 | 27.14 ± 0.08 |
| | B2 | **31.36 ± 0.07** | - | 30.94 ± 0.08 | 28.90 ± 0.10 | 29.31 ± 0.08 |
| NRMSE | K1 | **0.140 ± 0.001** | 0.397 ± 0.004 | 0.157 ± 0.001 | 0.177 ± 0.001 | 0.170 ± 0.001 |
| | K2 | **0.134 ± 0.001** | 0.376 ± 0.004 | 0.155 ± 0.001 | 0.174 ± 0.001 | 0.150 ± 0.001 |
| | K3 | **0.112 ± 0.001** | 0.317 ± 0.003 | 0.124 ± 0.001 | 0.135 ± 0.001 | 0.116 ± 0.001 |
| | B1 | **0.158 ± 0.002** | - | **0.159 ± 0.002** | 0.201 ± 0.002 | 0.240 ± 0.003 |
| | B2 | **0.162 ± 0.002** | - | 0.171 ± 0.002 | 0.218 ± 0.003 | 0.205 ± 0.002 |

# Results

## Knee Experiments

In our experiments involving knee MR images, we treated the highly under-sampled PDFSw *k*-space measurements as the "main information" and reconstructed a PDFSw RSS image from them, using the corresponding moderately under-sampled PDw *k*-space measurements (which we treated as "side information"). To evaluate TGVN's effectiveness in diverse settings, we conducted three experiments—K1, K2, and K3—with different sampling rates in main and side information, encompassing both the non-trivial and the trivial null space cases (i.e., acceleration factors greater than and less than the number of coils), respectively. Specifically, in experiment K1, the null space is nontrivial, enabling methods that exploit range-null space decomposition to be effective. In contrast, in experiments K2 and K3, the null space is trivial (i.e., contains only the zero vector), in which case such methods are expected to fail, as there is no meaningful null-space component to leverage.

**Q1:** Fig. 2 (top) shows the reconstruction results for coronal PDFS images with and without using the side information. At 20 × acceleration, side information significantly aids reconstruction while its absence results in loss of fine anatomical details highlighted by the yellow arrows. Table 2 further supports this finding, showing that TGVN consistently achieves superior performance across all evaluation metrics compared to E2E-VarNet.

Q2: Fig. 3 compares TGVN reconstructions against several baselines using side information and Fig. 4 illustrates three additional close-up cases in which TGVN demonstrates superior recovery of clinically relevant features—namely, meniscus tears and a meniscal cyst——compared to the baselines. MTrans and MC-VarNet exhibit significant blurring of anatomical features, and DMSI suffers severely from noise amplification, which is seen clearly in the absolute difference images. The output of TGVN is significantly superior: both overall sharpness and assorted anatomical details are better preserved in the TGVN reconstructions. Furthermore, the meniscus tears are distinctly more noticeable with TGVN, highlighting that it is *more effective* in leveraging the side information to preserve key features in the image despite highly sparse measurements. Notably, the meniscus tears are not well visualized in the images corresponding to the side information, which demonstrates that the TGVN is not relying excessively on features copied directly from the side information. Table 2 presents quantitative results, showing that TGVN achieves the best performance across all metrics with statistically significant improvements. In each experiment and for each evaluation metric (SSIM, PSNR, and NRMSE), a Wilcoxon signed-rank test rejected the null hypothesis at a significance level of 5%, indicating a statistically significant difference, and TGVN outperformed the next-best baseline in almost all test slices.

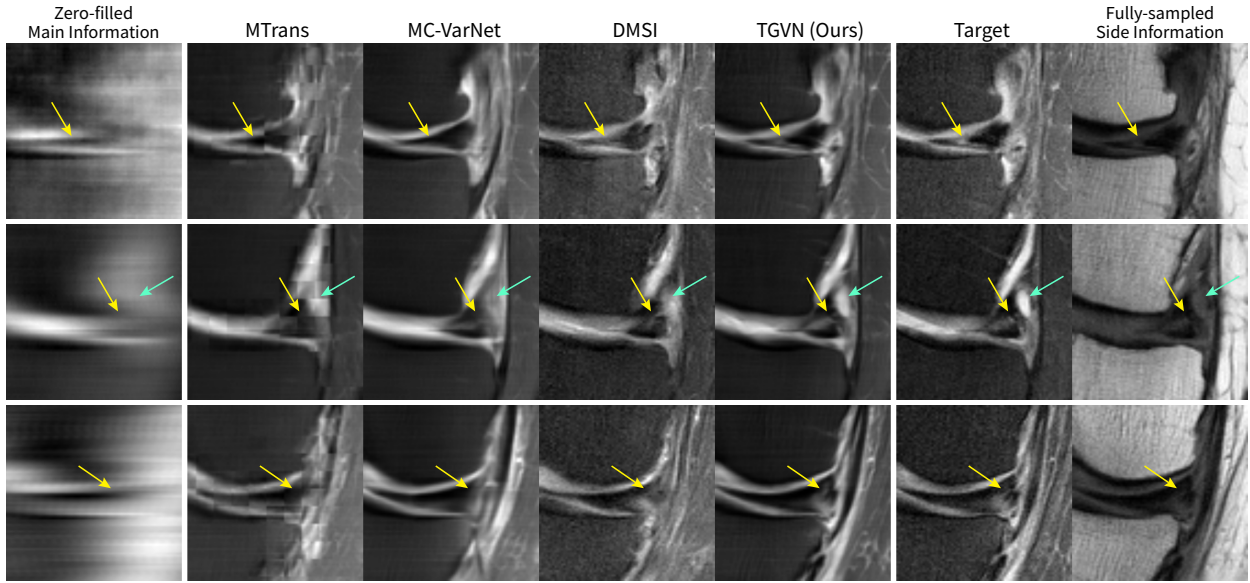

**Figure 4:** *Visualization of pathologies from K1 showing the effectiveness of TGVN in leveraging side information. In all three zoomed-in examples, meniscus tears (yellow arrows) are clearly visible only in the TGVN reconstruction, despite being significantly less visible in even fully sampled side information. The same is true for a meniscal cyst (green arrows) in the middle row.* **From left to right:** *zero-filled reconstruction; reconstructions from various baselines utilizing side information; TGVN reconstruction; target image; and fully sampled side information image. Models had access only to $2\times$ under-sampled side information in K1. Fully sampled side information is shown for illustration only.*

## Brain Experiments

In our experiments B1 and B2 using brain MR images, we used highly under-sampled FLAIR *k*-space and T1w *k*-space measurements from a single repetition as the "main information," aiming to generate reconstructions with image quality approaching that of signal-averaged multi-repetition

FLAIR and T1w RSS images, respectively. In both B1 and B2, the corresponding low-SNR, single-repetition T2-weighted (T2w) *k*-space measurements were used as "side information." Note that the protocol includes two repetitions for FLAIR and three for T1w and T2w. Hence, using a single repetition as side information, we achieved a practical acceleration factor of 3 ×. To demonstrate robustness to undersampling patterns, we used an equispaced mask in these experiments, providing a complementary setting to the random undersampling used in the knee experiments. Fully sampled T2w images were used as side information, in light of the low SNR and small matrix size of the acquisition [49].

**Q1:** Fig. 2 (bottom) shows the reconstruction results for axial FLAIR images with and without using the side information. At 36 × practical acceleration, side information aids the reconstruction significantly while reconstruction without side information results in the loss of various essential features. Moreover, Table 2 confirms that TGVN outperforms E2E-VarNet across all evaluation metrics.

**Q2:** Fig. 5 juxtaposes axial T1w images reconstructed using TGVN with corresponding images reconstructed using baseline methods. Use of side information results in substantial improvements in image quality at the challenging practical acceleration level of 45 ×. TGVN demonstrates superior performance in integrating this information compared to other methods, as evidenced by the enhanced depiction of anatomical features in the zoomed-in region and the consistently improved reconstruction metrics. Furthermore, for each quantitative evaluation metric (SSIM, PSNR, and NRMSE), a Wilcoxon signed-rank test rejected the null hypothesis at a significance level of 5%, indicating that there is a statistically significant difference between $s_{\text{ours}}$ and $s_{\text{base}}$. Table 2 reports the quantitative evaluation results for TGVN compared to baselines. As is the case for knee experiments, TGVN again achieves the best average score across all metrics. The statistically significant performance difference between TGVN and baseline methods indicates that the side information is beneficial in guiding the reconstruction, and that TGVN is more effective than other methods in leveraging it.

## Ablation Studies

We conducted ablation studies to address Q3 (whether projecting onto the *ambiguous space* provides benefits compared to no projection) and Q4 (how robust TGVN is to different undersampling factors, misregistration between main and side information, and degradation in side information quality). These studies also reinforce our main findings in Secs. 5.1 and 5.2.

## Effect of Projection

To answer Q3, we compared our proposed method with and without the projection. In particular, we compared the reconstruction performance of the unrolled network implementing (5) and the unrolled network implementing a modified version of (5) in which $\mathcal{P}_\delta$ is replaced by the identity operator. Starting with $\mathbf{x}_0 = \mathcal{A}^H \mathbf{k}$, the network without the projection implements the following update equations for $T$ iterations:

$$\mathbf{x}^{t+1} = \mathbf{x}^t - \eta^t \mathcal{A}^H(\mathcal{A}\,\mathbf{x}^t - \mathbf{k}) - \mu^t\bigl(\mathbf{x}^t - \mathcal{H}(\mathbf{s}; \gamma^t)\bigr) - \Phi(\mathbf{x}^t; \theta^t). \tag{8}$$

We applied an overall 9 × undersampling mask with equispaced 15 × outer undersampling and a 4% fully sampled center to the FLAIR measurements from a single repetition, achieving a practical acceleration factor of 18 ×. As in K1, fully sampled T2w images were used as side information. We performed three Wilcoxon signed-rank tests, and they rejected the null hypotheses at a significance level of 5%, concluding that there is a statistically significant difference between $s_{w/}$ and $s_{w/o}$, where $s_{w/}$ and $s_{w/o}$ represent the SSIM, PSNR, and NRMSE scores calculated on the test dataset for the TGVN with and without the proposed projector, respectively. Furthermore, for each metric, a scatter plot comparison of $s_{w/}$ vs $s_{w/o}$ demonstrates that the projection improves reconstruction quality for almost all slices in the test dataset.

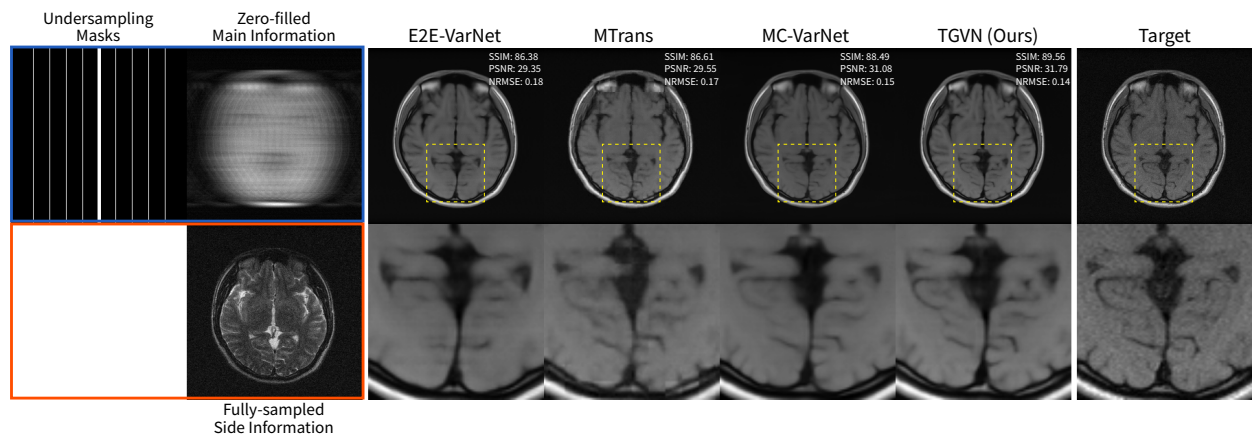

**Figure 5:** *Brain image reconstructions from B2 showing the effectiveness of TGVN at the challenging acceleration level of $45\times$ ($15\times$ undersampling and $3\times$ repetition reduction), in comparison to baselines* (DMSI omitted since it does not handle repetition reduction). **Top:** *Main information undersampling mask and full field of view images.* **Bottom:** *Fully sampled side information and zoomed-in regions indicated by dashed yellow boxes in the top row. TGVN preserves neuroanatomical structures and fine textures better than other methods.*

## Robustness to Degraded Side Information

Q4 was answered by conducting multiple experiments covering misregistered, undersampled, and irrelevant side information.

*Misregistration:* We compared the performance of TGVN when the side information is perfectly registered with performance in the presence of random misregistrations simulated by small random shifts and rotations during training and/or inference. We applied the same $9\times$ undersampling / $18\times$ acceleration scheme as in Sec. 5.3.1. For each slice, three random variables, $dx$, $dy$, and $d\theta$, were drawn uniformly from the interval $[-4, 4]$, and side information was translated by $dx$ and $dy$ pixels and rotated by $d\theta$ degrees. As expected, we observed that if TGVN does not encounter misregistration during training, the performance degrades substantially during inference. However, data augmentation with small random misregistrations during training renders the TGVN robust to small misregistrations during inference, as seen in Fig. 6. With such augmentation, TGVN still achieves significantly better scores than E2E-VarNet of the same capacity. This observation is supported by Wilcoxon tests at a 5% significance level for each metric—SSIM, PSNR, and NRMSE—demonstrating a statistically significant performance differences in favor of TGVN with misregistered side information, as compared to E2E-VarNet, which does not utilize side information.

*Undersampling:* We compared the performance of TGVN and MC-VarNet at four main and three side acceleration levels, and E2E-VarNet at these four main acceleration levels using the knee data. That is, we conducted experiments with all pairs of main acceleration factors (6 ×, 10 ×, 14 ×, and 20 × random undersampling) and side acceleration factors (2 ×, 3 ×, and 5 × uniform undersampling), for both TGVN and MC-VarNet (the second-best-performing method in K1 and K2). Comparable-capacity E2E-VarNet models were also trained at these main acceleration factors to facilitate further comparison. In Fig. 7, we plot the mean SSIM with standard errors as a function of the main and side acceleration factors. TGVN is clearly more robust to degraded (i.e., under-sampled) side information at all acceleration factors, while MC-VarNet with 5 × accelerated side information achieves performance comparable to E2E-VarNet (which utilizes no side information). Furthermore, when varying the main acceleration, there is a consistent performance gap between TGVN and MC-VarNet at each acceleration level, with TGVN performing better even at moderate main acceleration rates (e.g., 6 ×, 10 ×). As expected, the benefit of incorporating side information becomes more pronounced at higher acceleration levels (e.g., 14 ×, 20 ×).

In the second part of this ablation study, we investigated the effect of fully sampled side information, as well as a two-stage approach where the under-sampled side information was first reconstructed before it was utilized in reconstructing the main information. That is, we compared three models: (I) TGVN utilizing under-sampled side information; (II) TGVN utilizing under-sampled side information reconstructed first with an E2E-VarNet; and (III) TGVN utilizing fully sampled side information. We note that the TGVNs in (I), (II), and (III) have the same number of parameters; however, inclusion of the additional high-capacity E2E-VarNet in (II) introduces a minor element of unfairness in the comparison. Specifically, we conducted an experiment using knee data (K2) to compare TGVN performance using unmodified under-sampled side information $\mathbf{s}$ to that using independently-reconstructed under-sampled side information or fully sampled side information $\tilde{\mathbf{s}}$. Unmodified PDw under-sampled side information was provided to one TGVN (I), while fully sampled PDw side information was provided to another TGVN (III). For a third TGVN (II), a two-stage reconstruction was used, in which the 3 × under-sampled side information was first reconstructed with an E2E-VarNet $\zeta$ with 30 million trainable parameters, and the reconstructed result was then utilized as side information for the TGVN. As $I(\mathbf{x}; \tilde{\mathbf{s}}|\mathbf{k}) > I(\mathbf{x}; \mathbf{s} \mid \mathbf{k})$ for the target image $\mathbf{x}$ and the undersampled $k$-space $\mathbf{k}$, one would expect (III) to outperform (I). Moreover, since $\mathbf{x} \to \mathbf{s} \to \zeta(\mathbf{s})$ form a Markov chain, the data processing inequality [41] implies that (I) should perform at least as well as (II) in the infinite data regime. We observed that the reconstruction scores improve with fully sampled side information, but the improvements are slight and difficult to appreciate visually. Furthermore, despite having 30 million more trainable parameters to work with, the two-stage approach did not provide statistically significant improvements compared to TGVN utilizing under-sampled side information. Quantitative evaluation results for this experiment are provided in Table 3. Our takeaway from this experiment is that while fully sampled side information provides the greatest benefit, moderately under-sampled side information is still helpful, significantly improving the reconstruction compared to not having any side information (cf. Table 2, E2E-VarNet column). Furthermore, end-to-end training with under-sampled side information performs as well as the two-stage approach—a demonstration of the data processing inequality in the finite-data regime.

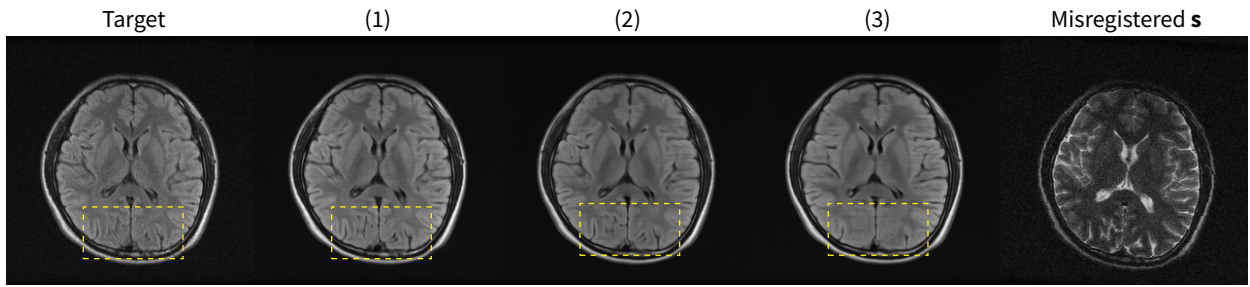

**Figure 6:** *Representative reconstructed images for Ablation Study 5.3.2.1. **Left:** Target image. (1): Reconstruction using TGVN trained with registered side information, encountering registered side information during inference. (2): Reconstruction using TGVN trained with augmentations simulating misregistrations, encountering misregistered side information during inference. (3): Reconstruction from E2E-VarNet without access to side information. **Right:** Misregistered side information. The most prominent differences are located inside the dashed yellow boxes. **Despite randomly misregistered side information, (2) preserves anatomical details much better than (3).***

*Relevance:* We compared the performance of TGVN using the worst possible side information (i.e., complex Gaussian noise) to that of E2E-VarNet with the same refinement block capacity, and also to that of TGVN using PDw side information. We trained and tested three models with the same refinement block capacity at 20 × random undersampling as in K1: (1) TGVN with 2 × undersampled PDw side information, (2) TGVN with random complex side information (Gaussian distributed in real and imaginary channels, with zero mean and a covariance matrix matching that of the side information in (1) in order to ensure that the side information amplitude is comparable to that of the main information), and (3) E2E-VarNet. To interpret how much the model relies on side information, we trained each of these models using only a single cascade element. This setup avoids ambiguity in interpreting learned parameters across multiple cascade elements, as both the trust-guidance coefficient $\mu$ and the data-consistency coefficient $\eta$ are allowed to be different for each element in a multi-element cascade architecture. With a single cascade, the ratio $\mu/\eta$ provides a clean proxy for the model's reliance on side information relative to the measured $k$-space. Since $\mu$ weights the trust guidance term and $\eta$ weights the data consistency term, this ratio reflects the model's internal assessment of the conditional mutual information $I(\mathbf{s}; \mathbf{x}|\mathbf{k})$ during training. For the model with the best validation SSIM, the ratio $\mu/\eta$ in (1) is 0.918, while for random side information in (2) it is 0.032, clearly indicating that TGVN learns to ignore side information when it is irrelevant. We report the quantitative evaluation results in Table 4, demonstrating that there is no statistically significant difference between a TGVN with irrelevant side information (2) and an E2E-VarNet (3), provided that the irrelevant side information—without any distribution shift—is present during both training and inference.

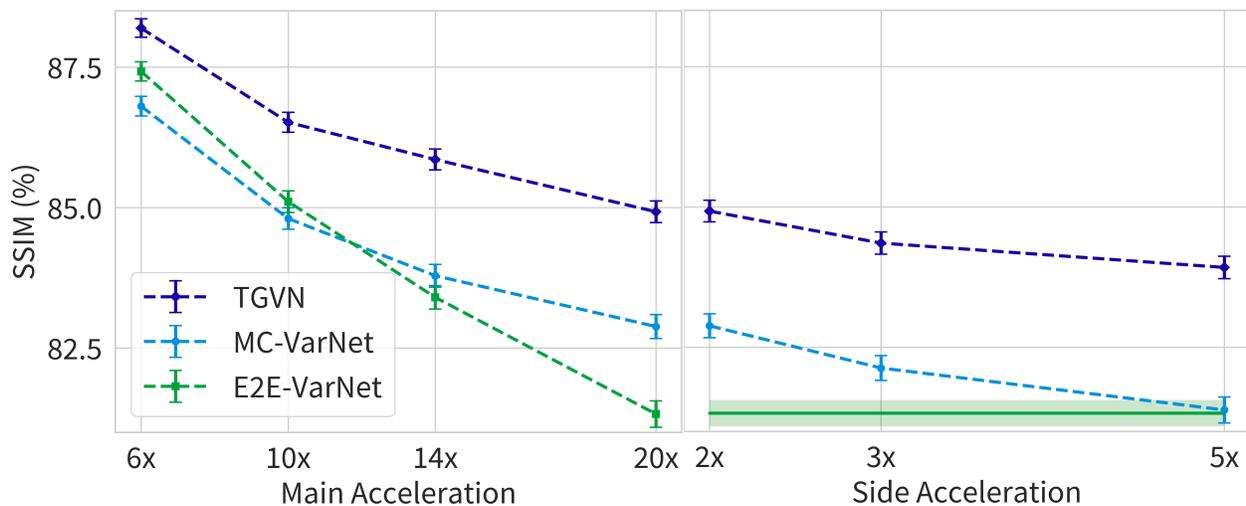

**Figure 7:** *Results from the first part of the Ablation Study 5.3.2.2 – Mean SSIM (%) as a function of main and side acceleration factor for TGVN, MC-VarNet and E2E-VarNet.* Left: Mean SSIM vs. main acceleration, with side acceleration at 2 ×. Right: Mean SSIM vs. side acceleration, with main acceleration at 20 ×.

**Table 3:** *Quantitative evaluation results for the second part of the Ablation Study 5.3.2.2.*

| Metric/Model | (I) | (II) | (III) |
|---|---|---|---|
| SSIM (%) | $85.52 \pm 0.19$ | $85.56 \pm 0.19$ | $\mathbf{85.97 \pm 0.18}$ |
| PSNR (dB) | $31.31 \pm 0.07$ | $31.41 \pm 0.07$ | $\mathbf{31.60 \pm 0.07}$ |
| NRMSE | $0.13 \pm 0.001$ | $0.13 \pm 0.001$ | $0.13 \pm 0.001$ |

**Table 4:** *Quantitative evaluation results for the Ablation Study Relevance 5.3.2.3.*

| Metric/Model | (1) | (2) | (3) |
|---|---|---|---|
| SSIM (%) | $\mathbf{83.75 \pm 0.20}$ | $80.15 \pm 0.24$ | $80.19 \pm 0.24$ |
| PSNR (dB) | $\mathbf{30.03 \pm 0.07}$ | $28.47 \pm 0.07$ | $28.47 \pm 0.07$ |
| NRMSE | $\mathbf{0.16 \pm 0.001}$ | $0.19 \pm 0.001$ | $0.19 \pm 0.001$ |

# Conclusion

We introduced a novel framework, the **T**rust-**G**uided **V**ariational **N**etwork (**TGVN**), that demonstrates the power of leveraging side information in solving LIPs, with specific application to the MR image reconstruction problem. By learning to eliminate solutions from the *ambiguous space* of the forward operator while remaining faithful to acquired measurements through *data consistency*, our principled approach makes maximal use of relevant side information while minimizing the risk of hallucinations. Our key finding is that, when incorporated effectively, subject-specific side information can significantly improve reconstruction quality and preserve anatomical and pathological features while avoiding hallucinations, even at undersampling levels as high as 20 ×, and even with moderately-under-sampled or low-quality side information. High levels of acceleration, meanwhile, can have a transformative impact in healthcare, by improving imaging efficiency in traditional settings, and also by enabling use of lower-quality data from accessible imaging devices for widespread health monitoring at the population level.

We performed a number of ablation studies in order to assess both the value of trust guidance and the residual risk of hallucinations when side information is degraded in quality or poorly matched to the main information. Our ablation results (Sec. 5.3) demonstrate that projection onto the ambiguous space does indeed provide notable performance benefits. They also show that on average, as long as training is adapted and there is no severe distribution shift, *side information cannot hurt the TGVN reconstruction*. In the worst case—when $\mathbf{s}$ and $\mathbf{x}$ are conditionally independent given $\mathbf{k}$—TGVN learns to ignore the side information by shrinking the trust-guidance coefficient $\mu^t$.

## Limitations and Future Work

In evaluating the effects of side information of varying quality, we have only considered a subset of possible image quality degradations. A wide range of practical imaging artifacts (susceptibility artifacts, through-plane and non-rigid motion, effects of eddy currents and gradient nonlinearities, etc.) can result in mismatches between main and side information. Our training set did naturally include some instances of these artifacts, but not in sufficient numbers to explore thoroughly the space of possible mismatches. While we expect that our principal findings regarding the robustness of TGVN performance will be preserved in most of these settings, this hypothesis remains to be tested, and further evaluations are planned as future work.

Given the promising results of TGVN reconstruction using complementary-contrast measurements from the same MR examination as side information, we intend to expand our studies to explore incorporation of different types of side information. When the side information arises from a cross-sectional imaging modality (MR, PET, CT), the $\mathcal{H}$ block can continue to be a U-Net. If the side information takes the form of text, it can first be encoded with a pre-trained text encoder and the $\mathcal{H}$ block must then decode the resulting representation back to the image domain. Essentially, $\mathcal{H}$ is a learnable transformation from the domain of the side information to the domain of the complex-valued, coil-combined MR image such that the difference $\mathbf{x} - \mathcal{H}(\mathbf{s}; \gamma)$ in the trust-guidance term is well-defined. Future work will involve use of a patient's prior scans and associated textual data (e.g., clinical notes and medical history) as side information. We will also explore the potential value of using features learned from related tasks to inform trust-guided image reconstruction.

## Appendix

## Training and Evaluation Details

### TGVN

We used the MS-SSIM-L1 [50], [51] loss function and employed the ADAM optimizer [52], with batch size of one per GPU and with default parameters including a uniform kernel of size $33 \times 33$ and $k$-values of 0.01 and 0.03, across 5 values of $\sigma$ (0.5,1.0,2.0,4.0,8.0), in both training and validation phases. The loss is calculated between the reconstructed and the ground-truth root-sum-of-squares (RSS) [53] images. A starting learning rate of $3 \times 10^{-4}$ was used with exponential decay with the decay parameter 0.98. These parameters were determined through a grid search on the validation set. The training spanned 100 epochs, with the best model parameters selected based on the validation loss. All models were trained and tested on 4 × NVIDIA A100 GPUs using PyTorch for 10 days in knee experiments and 2 days in brain experiments. Sensitivity maps are estimated with a separate module (SME), as in [6]. The ASC ($\mathcal{H}$), refinement ($\Phi$), and SME modules are all implemented using a U-Net model, as illustrated in Fig. 1, which consists of four down-sampling and up-sampling paths, complemented by skip connections. Each path is equipped with two 3 × 3 convolutions, followed by instance normalization [54] and leaky ReLU [55] activation functions. The first convolution layer outputs 21 channels in the ASC and refinement networks and 8 in the SME network, with the number doubling in each subsequent layer, as in [6]. In the brain experiments, the proposed TGVN comprises $T = 10$ TGVN blocks having approximately 67.3 million trainable parameters in total. In the knee experiments, it includes $T = 14$ TGVN blocks, resulting in about 94 million trainable parameters. For enhanced numerical stability in training, complex-valued layer normalization and its inverse operator are used, similar to [6]. Specifically, each network pass, denoted as $\mathbf{x} \mapsto \Phi(\mathbf{x})$, is executed as $\mathbf{x} \mapsto \mathcal{T}^{-1}(\Phi(\mathcal{T}\mathbf{x}))$. For an input tensor of shape (B, 2, H, W), where $B$ is the batch size, the two channels correspond to the real and imaginary components of the image. The normalization process $\mathcal{T}$ adjusts each sample so that the mean of each channel across the spatial dimensions $(H, W)$ is zero, and the standard deviation is set to one. Furthermore, $\mathcal{T}$ ensures that the real and imaginary channels are decorrelated, resulting in zero covariance between them. This is achieved by computing the 2 × 2 covariance matrix of the two channels and performing a linear combination with the transpose of the Cholesky decomposition of the inverse of the covariance matrix. This normalization step allows the network to handle the real and imaginary parts without any inherent bias or unintended correlation, ultimately improving the stability and performance of the model. In both knee and brain experiments, we implemented the approximate projector as described in Sec. 3.2 for the trust-guidance term with 10 Conjugate Gradient iterations, and learned the threshold parameter $\delta$ during training. Use of the approximate projector circumvented the need for a computationally expensive SVD operation. For example, in K3, the forward operator has more than 200,000 rows and 200,000 columns, making explicit SVD calculation prohibitively costly.

## Baselines

For the baselines, we used the officially released repositories instead of re-implementing the models. As a result, we only needed to adjust the model capacities to match that of TGVN in each setting and modify the training learning rates to adapt to the fastMRI and M4Raw datasets.

*E2E-VarNet:* We used the MS-SSIM-L1 loss function and employed the ADAM optimizer, with batch size of one per GPU and with the same hyperparameters as used in TGVN. The loss is calculated between the reconstructed and the ground-truth root-sum-of-squares (RSS) images. A starting learning rate of $3 \times 10^{-4}$ was used with exponential decay with parameters 0.98. These parameters were determined through a grid search on the validation set. The training spanned 100 epochs, with the best model parameters selected based on the validation loss. All the models were trained and tested on 4 × NVIDIA A100 GPUs using PyTorch for 3 days in knee experiments and 1 day in brain experiments. The refinement and SME modules share the same U-Net architecture as TGVN. The first convolution layer outputs 30 channels in the refinement network and 8 in the SME network, with the number doubling in each subsequent layer, as in [6]. In the brain experiments, it comprises $T = 10$ VarNet blocks having approximately 68.7 million trainable parameters in total. In the knee experiments, it includes $T = 14$ VarNet blocks, resulting in about 96 million trainable parameters.

*MTrans:* We used the CrossCMMT model and set the hidden dimension to 17 and 14 for the knee and brain experiments, respectively. Input sizes are chosen as the image matrix size, without any resolution change. Parameters P1, P2 are set to 8, and `CTDEPTH` and `TRANSFORMER_NUM_HEADS` are set to 4, and `TRANSFORMER_MLP_RATIO` is set to 3. With these parameters, the knee model has 98.3 million and the brain model 66.1 million trainable parameters. Initial learning rate was set to $2 \times 10^{-4}$. We trained one model without scheduling and one model scheduled with an exponential decay $\gamma = 0.99$. Both models were trained for 100 epochs on 4 × NVIDIA A100 GPUs using PyTorch with a unit batch size per GPU. The best-performing model, according to the average validation SSIM, was the one trained with exponential decay. Training spanned approximately 4 days for knee and 1 day for brain.

*MC-VarNet:* We set the `in_channel` and `channel_fea` parameters to 264 and 224 without changing the default `iter_num` (4), which result in 94.1 and 67.8 million trainable parameters for the knee and brain experiments, respectively. For the brain experiments, we used an initial learning rate of $10^{-4}$, and trained one model without scheduling and one model scheduled with an exponential decay parameter 0.99. For the knee experiments, learning rates on the order of $10^{-4}$ resulted in unstable training, so we chose an initial learning rate of $10^{-5}$ and again trained one model without scheduling and one with exponential decay parameter 0.99. Each model was trained for 100 epochs on 4 × NVIDIA A100 GPUs using PyTorch with a unit batch size per GPU. The best-performing model, according to the average validation SSIM, was the one trained with exponential decay. Training spanned approximately 7 days for knee and 1 day for brain.

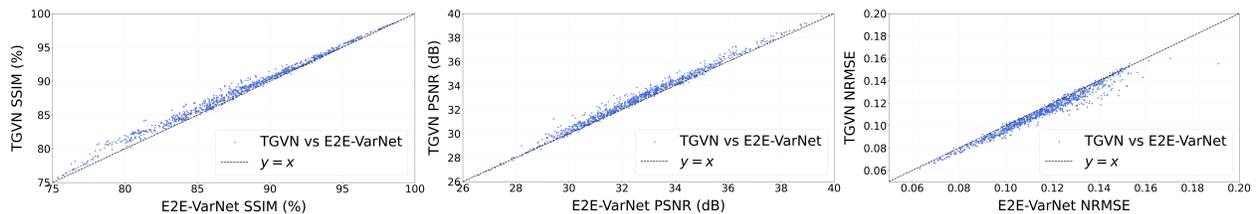

**Figure 8:** *Quantitative evaluation results for SSIM, PSNR, and NRMSE over the test dataset for K3.*

*DMSI:* By design, DMSI reconstructs complex-valued coil-combined images. From the reconstruction $\hat{\mathbf{x}}_{\text{DMSI}}$, we obtained the reconstructed RSS combination using $\sqrt{\sum_i |S_i\, \hat{\mathbf{x}}_{\text{DMSI}}|^2}$ and compared against ground-truth RSS images. (This gave better performance than comparing $|\mathbf{x}_{\text{DMSI}}|$ with the ground-truth RSS images.) To train the score network, we used SongUnet network architecture with positional embedding, and standard encoder and decoder. We used default parameters in the codebase, only changing the `model_channels` to 210 to match the trainable number of parameters, resulting in approximately 92.4 million parameters. We employed augmentation with p = 0.12 and dropout with p = 0.13. We trained the model to minimize the EDM loss for 400,000 steps with a batch size of 4 per GPU. The training spanned approximately 10 days.

## Evaluation

For the SSIM metric, a 7 × 7 uniform kernel was utilized, along with the standard $k-$values of 0.01 and 0.03. The range parameter was given as input to the SSIM calculation and was set to the maximum pixel value of the corresponding volume. In Wilcoxon signed rank tests, we let $\mathcal{D}$ be the distribution of pairwise differences $s_{\text{ours}} - s_{\text{base}}$. Then under the alternate hypothesis, $\mathcal{D}$ is "stochastically greater than a distribution symmetric about zero" for SSIM and PSNR and

"stochastically less than a distribution symmetric about zero" for NRMSE. The average inference time on an NVIDIA A100 GPU for TGVN was measured to be 468 ms per slice. In comparison, the reconstruction times per slice were 132 ms for E2E-VarNet, 43 ms for MTrans, 495 ms for MC-VarNet, and 199,206 ms for DMSI—more than 400 × slower than the second-slowest method, MC-VarNet.

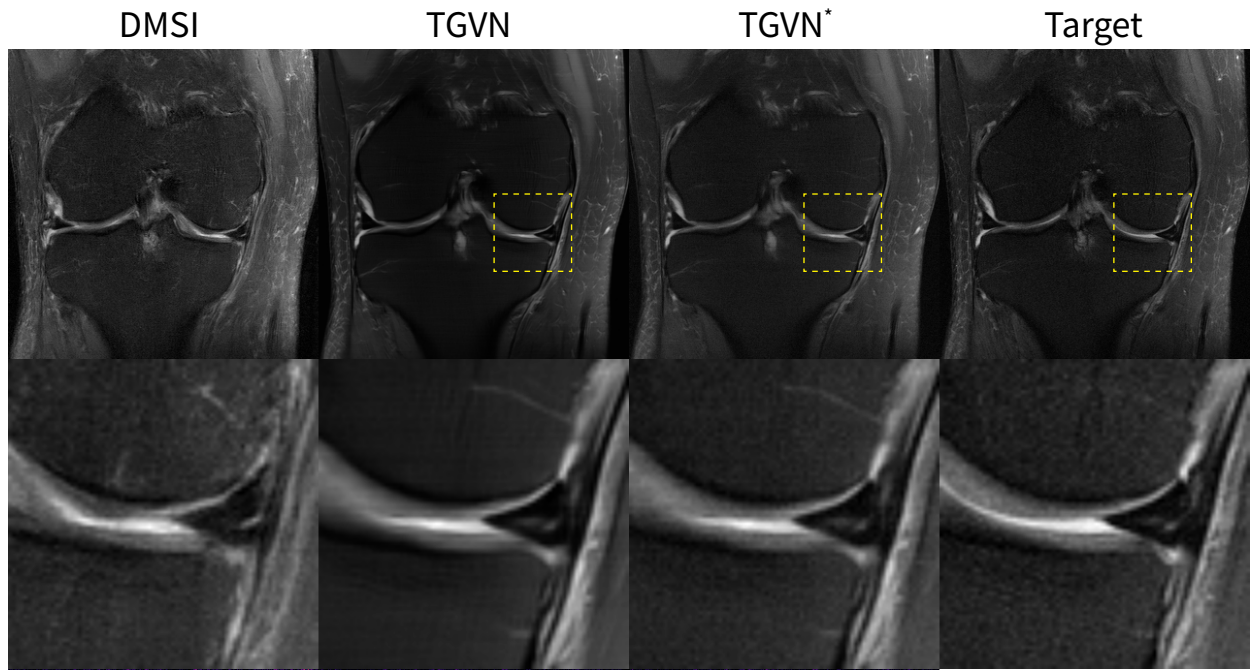

**Figure 9:** *Selected knee image reconstructions from K1, including TGVN with added image noise. A common feature of end-to-end reconstructions compared to generative approaches such as DMSI is a residual denoising effect, which can give the appearance of smoothing fine structures or background textures. To enhance the subjective perception of sharpness in images, known as acutance in photography, low levels of Gaussian noise were added back to the reconstructed TGVN output, which is represented as TGVN\*. Improved preservation of key features may be appreciated in the TGVN\* reconstruction as compared with the DMSI reconstruction.*

## Additional Evaluation Results

Scatter plots comparing SSIM, PSNR, and NRMSE scores on the test split for TGVN and the second-best method for K1–3, B1–2 demonstrate that TGVN consistently improves reconstruction quality for almost all slices in the test dataset. One example is provided in Fig. 8. The remaining examples are not shown due to space constraints. To enhance visual comparison despite the TGVN's inherent denoising properties, we performed additional reconstructions shown in Fig. 9 with noise added deliberately to match the subjective noise level of the target images, as previously implemented in [56]. This preserved not only anatomical and pathological features but also overall image appearance as compared with target images.

## Acknowledgment

The authors thank Michael P. Recht, MD, for his guidance regarding pathological features and image quality in Figs. 3 and 4.